\documentclass[superscriptaddress,twocolumn,showpacs,preprintnumbers,amsmath,amssymb,prb]{revtex4-1}
\usepackage{graphicx,times,epsfig,color}

\usepackage{graphicx,times,color}
\usepackage{psfrag}
\usepackage{epsfig}
\usepackage{verbatim}

\begin{document}

\title {Transport properties of a two impurity system: a theoretical approach.}

\author{I. J. Hamad}
\affiliation{Departamento de F\'{\i}sica, Pontif\'{\i}cia Universidade Cat\'olica do Rio de Janeiro, 22453-900, Brazil}
\affiliation{Instituto de F\'{\i}sica Rosario, Universidad Nacional de Rosario, Bv. 27 de Febrero 210 bis, Rosario 2000, Argentina}
\email[Corresponding author: ]{hamad@ifir-conicet.gov.ar}

\author{L. Costa Ribeiro}
\affiliation{Centro Federal de Educa\c{c}\~ao Tecnol\'ogica Celso Suckow da Fonseca (CEFET-RJ/UnED-NI), RJ, Brazil}
\author{G. B. Martins} 
\affiliation{Department of Physics, Oakland University, Rochester, MI 48309, USA}
\author{E.~V. Anda}
\affiliation{Departamento de F\'{\i}sica, Pontif\'{\i}cia Universidade Cat\'olica do Rio de Janeiro, 22453-900, Brazil}

\date{\today}

\begin{abstract}
A system of two interacting Cobalt atoms, at varying distances, has been studied in a recent Scanning Tunneling Microscope 
experiment by J. Bork {\it et al.}, Nature Physics {\bf 7}, 901 (2011). We propose a microscopic model that explains, 
for all the experimentally analyzed interatomic distances, the physics observed in these experiments. Our proposal is 
based on the two-impurity Anderson model, with the inclusion of a two-path geometry for charge transport. This many-body system is 
treated in the finite-U Slave Boson Mean Field Approximation and the Logarithmic-Discretization Embedded-Cluster Approximation. 
We physically characterize the different charge transport regimes of this system at various interatomic distances and show 
that, as in the experiments, the features observed in the transport properties depend on the presence of two 
impurities but also on the existence of two conducting channels for electron transport. We interpret the splitting 
observed in the conductance as the result of the hybridization of the two Kondo resonances associated to each impurity.

\end{abstract}
\pacs{73.23.Hk, 72.15.Qm, 73.63.Kv}
\maketitle

\section{Introduction}

Two interacting magnetic impurities in a bath of conducting electrons is one of the simplest strongly correlated systems 
with a rich phase diagram containing a Kondo regime region and a spin singlet state where the two impurities are locked 
into a dimer \cite{Jones88}. The possible quantum phase transition (QPT) between these two phases is dominated by a 
non-Fermi liquid quantum critical point (QCP). The parameter that drives the system through the phase diagram is the ratio $I/T_K^0$, where $I$ is the inter-impurity exchange interaction and $T_K^0$ is the Kondo temperature of the individual impurities (assumed to be identical). This system has received great attention both theoretically (see below) and experimentally \cite{Chen99,Jeong01,Craig04,Wahl07,Neel11}. Experiments performed in two-impurity systems have been able to carry it from the Kondo screened phase to the antiferromagnetic (AF) regime, but without achieving a precise control of the exchange interaction between the two impurities.\cite{Craig04,Jeong01} 

From the theoretical point of view, numerical renormalization group (NRG) \cite{Bulla08} calculations on 
the two-impurity Anderson model (TIAM) focused on the properties of its non-Fermi Liquid QCP \cite{Sakai92_1} and 
pointed out that the interimpurity hopping suppresses the critical transition \cite{Affleck92,Gan95}. Several theoretical 
methods were used to analyze the TIAM, as the slave-boson formalism \cite{Aono98,Aono01,Aguado00,Vernek06,Dong02,Lopez02,Ribeiro12},  the NRG \cite{Sakai92_1,Izmuda00}, the Embedded Cluster Approximation (ECA) \cite{Busser00}, and the noncrossing approximation \cite{Aguado03}. 
The results obtained confirmed the replacement of the critical transition by a crossover as a consequence of the broken  
even-odd parity symmetry. In addition, it was observed a splitting of the zero-bias anomaly in the differential conductance (dC) with the increase of the interdot hopping. This can be understood from the coherent superposition of the many-body Kondo states of each QD (forming bonding and antibonding combinations) \cite{Aono01,Aguado00,Busser00,Ribeiro12} or, 
alternatively, due to the `parity splitting' caused by the direct hoping between the impurities \cite{Sakai90,Sakai92_1,Sakai92_2}.

Recently, a remarkable experiment has been performed where a Cobalt (Co) atom, positioned at the tip of 
a Scanning Tunneling Microscope (STM), is continuously approached to another Co atom adsorbed 
on an Au(111) surface \cite{Bork11}.  The position of the STM tip was varied with sub-picometer 
(pm) accuracy and hence the ratio $I/T_K^0$ (as defined above) could, in principle, be modified almost continuously. 
The results in Ref.~\onlinecite{Bork11} indicated that the system stayed away from the neighborhood of the QCP, as a 
peak in the dC, expected to appear at the QPT~\cite{DeLeo04,Sela-affleck09}, 
was not observed. Starting from a larger interatomic distance, where electron transport occurs through tunneling, the 
dC initially showed a Fano antiresonance that, with decreasing distance, first narrowed and then evolved into a peak. This peak in the dC, upon further approaching the Co atoms, showed a splitting that was interpreted as consequence of an effective exchange interaction between the two magnetic impurities\cite{Lopez02,Simon05,Wahl07}. 
However, surprisingly enough, this splitting was observed at energy scales smaller than $ T_K^0$ (for details, see Fig.~6 in Ref.~\onlinecite{Bork11}), 
in contradiction with theoretical results for two-impurity models that establish a critical coupling $I^*$ greater 
than the Kondo temperature ($\approx 2 k_B T_K^0$) at which the splitting should appear \cite{Georges99,Jones87,Jones88,Lopez02,Simon05}. 
In addition (Fig.~6 in Ref.~\onlinecite{Bork11}), the QPT was washed away by a broad crossover region, the lower end of which 
roughly coincides with the appearance of the splitting in the dC. The origin of this wide crossover was explained by Bork {\it et al.} \cite{Bork11}
as coming from the strong direct coupling between the electrodes, i.e., the STM tip and the Au substrate themselves. 

The Fano antiresonance has been discussed in Ref.~\onlinecite{Bork11} within the context 
of a phenomenological model and not by solving a  microscopic Hamiltonian. Such analysis 
assumes the existence in the dC of two Fano anti-resonances of Kondo 
origin with a superposition between them. However, this treatment is not able to obtain 
neither the  single peak at the Fermi level nor its splitting when the distance between the Co 
atoms is reduced, as observed in the experiments. To provide an explanation for this last 
feature of the dC, a microscopic model was proposed and solved within the NRG formalism, \cite{Bork11} 
incorporating an indirect coupling between the Co atoms. It is then important to theoretically 
account for such an evolution of the dC by using a realistic and {\it single} microscopic model, capable of 
reproducing all the experimental features described above. This model provides, as well, evidence that a splitting 
in the dC is compatible with both impurities still being independently in the Kondo screened regime. This complete explanation 
is particularly important because a consistent characterization of the physics observed in such an 
experiment, as a function of distance (or equivalently, interaction between the impurities), is still 
lacking in the literature.

\begin{figure}
\includegraphics[scale=0.15]{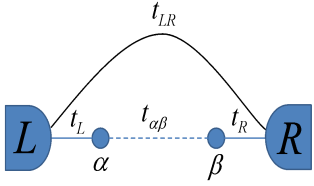}
\caption {(Color online) Schematics of the model studied in this work. $\alpha$ and $\beta$ represent the two Co atoms.}
\label{fig1}
\end{figure}

In this paper, we show that the double-Co experiment described above can be completely interpreted, 
for all inter-impurity distances studied, by a model that  incorporates, as essential ingredients,  
{\it a direct hopping between the Co atoms and another one between the electronic reservoirs}. 
In particular, the results show to what extent the interplay between the direct and indirect 
inter-impurity hoppings influences the transport properties of the system. Moreover, we show 
that the splitting in the conductance is compatible with a Kondo screened ground state.

\begin{figure}[ht]
\includegraphics[scale=0.5]{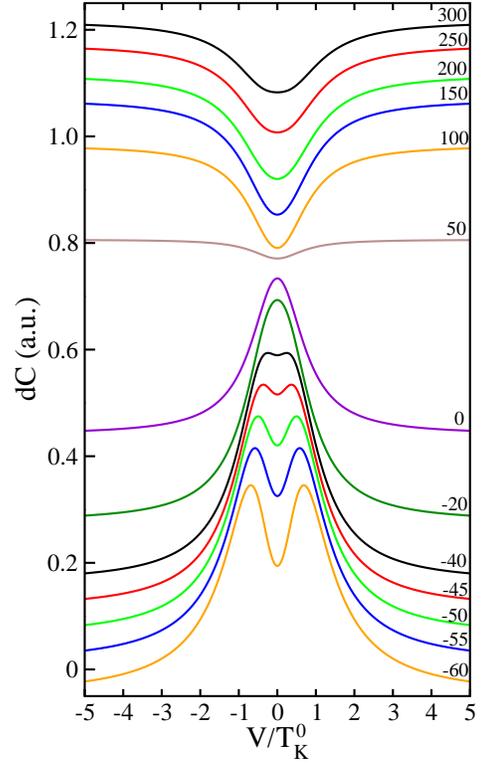}
\caption{(Color online) Differential conductance as a function of $V/T_K^0$ for $300 \leq z \leq -60$ 
(where $z$ is the separation between the impurities, measured in pm, indicated in the right side). Each curve, 
calculated using SBMFA, has been shifted vertically for clarity.}
\label{fig2}
\end{figure}

\section{Model}

The Hamiltonian is written as a sum of three terms, namely 
\begin{eqnarray}
H = H_{\rm {imp}}+H_{\rm {hyb}}+H_{\rm {leads}}, 
\end{eqnarray}
where 
\begin{eqnarray} 
H_{\rm {imp}}&=&\sum_{i=\alpha,\beta;\sigma}{\left(\epsilon_i n_{i\sigma}+\frac{U}{2} n_{i\sigma}n_{i\bar{\sigma}}\right)}
\end{eqnarray}
describes the isolated impurities, $\epsilon_{i}$ (where $i=\alpha,~\beta$) being the energy of each localized impurity, 
$U$ the on-site Coulomb interaction, and $\sigma=\pm$ is the spin orientation. The contribution
\begin{eqnarray}
H_{\rm hyb} &=& \sum_{\sigma} t_L c^{\dag}_{L,1\sigma} c_{\alpha\sigma}+t_Rc^{\dag}_{R,1\sigma}c_{\beta\sigma}+
t_{\alpha\beta}c^{\dag}_{\alpha\sigma}c_{\beta\sigma} \nonumber \\
&+& t_{LR}c^{\dag}_{L,1\sigma}c_{R,1\sigma} + \mbox{H.c.}
\end{eqnarray}
describes the hybridization of each impurity with the first site of its adjacent metallic lead, the hybridization between both 
impurities and the direct tunneling between the left (L) and right (R) electron reservoirs (in that order). Finally, 
\begin{eqnarray}
H_{\rm leads} &=&  t \sum_{j=L,R}\sum_{i=1;\sigma}^\infty \left(c_{j,i\sigma}^{\dagger} c_{j,i+1\sigma} +\mbox{H.c.}\right) ,
\end{eqnarray} describes the $L,R$ leads, represented by two semi-infinite chains of non-interacting sites, with hopping $t$ 
between adjacent sites. Note that the hopping between each impurity and the opposite electrode was not 
included, since the experiments indicated that these couplings do not play an important role in the charge transport, as no change 
in the dC line-shape for a bare tip approaching a Co atom was observed \cite{Bork11}. The interaction between 
each impurity and the opposite electrode is indirectly included through the hopping $t_{LR}$ between the 
leads. In addition, its explicit inclusion also proved unnecessary, as the experimental results could be explained with a 
simpler model. In fact, our results show that the hypothesis of Ref.~\onlinecite{Bork11} just mentioned is correct. 
The model is depicted in Fig.~\ref{fig1}. The transport properties were calculated within the finite-U Slave 
Boson Mean Field Approximation (SBMFA) \cite{sbmfa}, although, for the sake of comparison, some results were 
obtained using the  Logarithmic-Discretization Embedded-Cluster
Approximation (LDECA) \cite{Anda08}. Finally, for the sake of simplicity, we adopt a symmetric model
(i.e., $t_L=t_R=t^{\prime}$). 

We define $T_K^0$ as the Kondo temperature for each Co atom in the two independent single-impurity Anderson 
models (as obtained through $t_{LR}=t_{\alpha \beta}=0$). Taking, in units of $t$, $U=0.8$, $t^{\prime}= 0.25$, and $\epsilon_i=-U/2$, we 
obtain $T_K^0=0.0073$. Considering $t=1.3~{\rm eV}$ for Au, one obtains $T_K^0 \approx 9.5~{\rm meV} \equiv 110~{\rm K}$, 
roughly the same Kondo temperature measured in the experiments with Co atoms on Au \cite{Bork11}. $T_K^0$ will 
be a reference for comparison with experiments. The values of $U$ and $t^{\prime}$ were chosen, within the convergence 
parameter space of the SBMFA method, so that $U/\Gamma \sim 14$, where $\Gamma=\pi t^{\prime 2} \rho(E_F)$, 
being $\rho(E_F)$ the leads' DOS at the Fermi energy. This $U/\Gamma$ value assures that the single impurity system 
is deep inside the Kondo regime at zero temperature.  

\begin{figure}
\includegraphics[width=\linewidth]{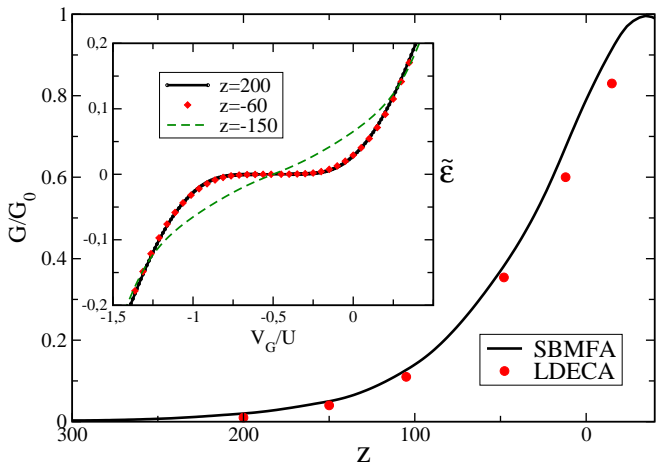}
\caption{(Color online) Main panel: Conductance $G/G_0$ as a function of $z$. When compared 
to Fig.~4(a) in Ref.~\onlinecite{Bork11}, we see the same overall behavior. Inset: renormalized energy 
level $\tilde \epsilon_{\alpha}=\tilde \epsilon_{\beta}=\tilde \epsilon$ as a function of gate voltage 
$V_g$. The plateau at the Fermi energy for both $z=200$ (continuous (black) line) and $z=-60$ [(red) diamonds line], 
indicates that the system stays in the Kondo regime. For larger hopping values, $t_{\alpha \beta}=0.06$ and $t_{LR}=0.13$, 
the plateau starts to be suppressed, as shown by the (green) dashed curve for $z=-150$, reflecting that the system enters a  
crossover regime (see text).}
\label{fig3}
\end{figure}

Our model includes the hopping $t_{LR}$ between the electron reservoirs, providing a channel through which 
the dots couple indirectly and also a weaker direct hopping $t_{\alpha \beta}$ between the Co atoms that, 
as mentioned above, results to be essential to reproduce the experimental results. The assumption that 
$t_{\alpha \beta} \ll t_{LR}$ is in accordance with the hypothesis made in Ref.~\onlinecite{Bork11}, namely, 
that for a vertical approach between the STM tip and surface, the interaction between the $d$-orbitals 
of the Co atoms is very weak. Assuming the hoppings to follow $t_{LR}=A e^{\gamma z}$ and 
$t_{\alpha \beta}=B e^{\delta z}$, where $z$ is a parameter representing the inter-impurity distance, 
then a decrease in $z$ results in an exponential increase in $t_{\alpha \beta}$ and $t_{LR}$ (for $\gamma,~\delta<0$). 
After an extensive survey, by varying the parameters so as to take into account the constraint $t_{\alpha \beta}<<t_{LR}<t$, 
as discussed above, and the obvious conditions  $t_{\alpha \beta} \approx 0$ and $t_{LR}<<1$ for the largest $z$-value used, 
we found  that the parameters that best allow the reproduction of the experimental results are \cite{note2} $A=0.4066$, $B=0.0305$, $\gamma=-0.002534$, and $\delta=-0.00973$. This parametrization allows us to reproduce the dC experimental results for the whole range of $z$ values (compare our Fig.~2 with Fig.~4(b) in Ref.~\onlinecite{Bork11}).
 
In the SBMFA, the dC is calculated using the Keldysh formalism \cite{Keldysh}. For simplicity, we assume electron-hole 
symmetry. The differential conductance can then be written as, 
$dC= 4\pi^2 t^{\prime 4} \mbox{Im}\{G_L(V/2)\} \mbox{Im}\{G_R(V/2)\} |G_{LR}^V(V/2)|^2 $, where $G_L=G_R$ are 
the reservoirs' non-interacting Green's functions and $G_{LR}^V(\omega)$ is the many-body propagator 
from L to R, under the presence of a bias $V$ between $L$ and $R$ reservoirs. 
However, it is known \cite{Lara} that for a 
two-impurity system the SBMFA results obtained for the equilibrium situation are very similar to the 
non-equilibrium results, as long as $V$ is smaller than a few times $T_K^0$. Under these conditions, the 
propagator $G_{LR}^V(\omega)$ is almost independent of the external bias $V$. Therefore, we will assume 
its complete independence from $V$ and calculate dC as if the system were in equilibrium.

\section{Results}

\begin{figure}
\includegraphics[width=\linewidth]{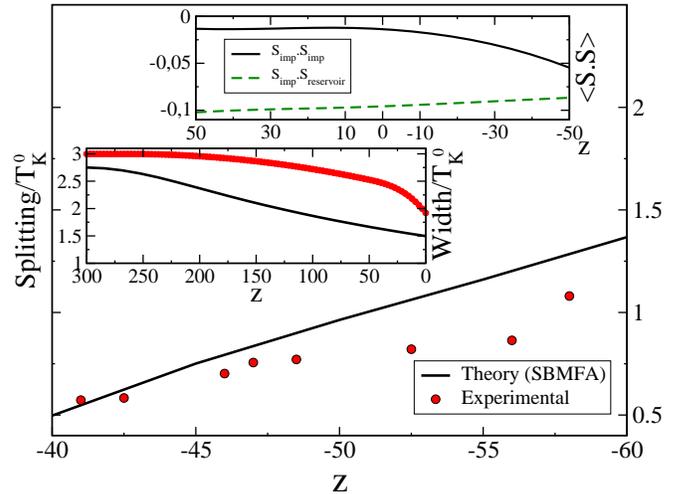}
\caption{Splitting as a function of $z$ [(black) solid curve]. Lower inset: Width of the 
anti-resonance, as a function of distance, in units of $T_K^0$. The SBMFA results (black line) are 
similar to those obtained in Figs.~5(b) and 5(a) in Ref.~\onlinecite{Bork11} [here reproduced 
schematically as solid (red) dots]. Upper inset: Spin-spin correlations between the impurities 
(black solid curve) and between each impurity and its adjacent reservoir (green dashed curve, see text).}
\label{fig4}
\end{figure}

In Fig.~\ref{fig2}, we present the SBMFA results for dC as a function 
of $V/T_K^0$. The dC curves present three fundamental features that should be emphasized: {\it (i)} for 
negative values of $z$, there is a double-peak structure, displaying a splitting that decreases with 
increasing $z$; {\it (ii)} eventually, still for negative values of $z$, the splitting is totally suppressed, 
becoming a single peak; {\it (iii)} for higher $z$ values, a Fano anti-resonance  develops centered at $V=0$, 
with increasing width as $z$ increases (see also Fig.~\ref{fig4}). The behavior just described is qualitatively and 
semiquantitatively similar to that observed in the experiments by Bork {\it et al.} (Fig.~4(b) of Ref.~\onlinecite{Bork11}). 
However, the range of $z$ values for which our results show a single-peak is slightly larger than in the experiments, 
where an anti-resonance line-shape persists down to $z \approx -30$, while our results show a single-peak feature already at $z=0$. 

Note that a split-peak in dC was also obtained by using Numerical Renormalization Group \cite{Bork11}, with 
a model where only the hopping between the electronic reservoirs was included. However, the overall agreement 
between our theoretical results and the experiments requires the inclusion, as a crucial parameter, of the 
{\it direct} inter-impurity interaction $t_{\alpha \beta}$. This direct hopping opens another channel 
through which electrons can flow, hence, at larger distances, this results in a Fano dip, when $t_{\alpha \beta}$ 
is very small. The dip transforms into a peak as the ratio $t_{\alpha \beta}/t_{LR}$ increases (from $0.009$ at $z=300$ 
to $0.115$ at $z=-60$). At the lower distances, (larger $t_{\alpha \beta}$) the Kondo states of each impurity superpose, 
forming bonding/antibonding many-body states that results on a splitting in the dC, as shown in our results in Fig.~\ref{fig2}. 
It is important to point out that a splitting (with varying magnitude in relation to $T_K^0$) in the dC has also been obtained in previous studies on similar models. In these studies, a direct hopping and/or a superexchange interaction J between 
the impurities has been taken in account \cite{Georges99,Sakai90,Sakai92_1,Sakai92_2,Aguado00,Vernek06,Busser00,Ribeiro12}. 
The splitting thus obtained was interpreted as a superposition of the 
two `independent' many-body Kondo states \cite{Aguado00,Vernek06,Busser00,Ribeiro12} or, alternatively 
(but not in contradiction with the previous idea), as caused by the parity splitting of the occupation number between 
even and odd channels, that takes place when the direct hopping between the impurities has a magnitude comparable to 
the low energy scale of the model, as in our case. \cite{Sakai90,Sakai92_1,Sakai92_2} 

Using SBMFA, we have calculated the phase difference (not shown) between the two channels through which the 
current can flow, the one between the impurities via $t_{\alpha \beta}$ and the one through 
the reservoirs $t_{LR}$. \cite{note2a} We obtained that this phase difference is zero for all values of $z$, 
indicating that these two channels interfere constructively, satisfying the Onsager relation for systems of closed geometry, which establishes that this phase difference can be only zero or $\pi$. 

In Fig.~\ref{fig3} we present SBMFA (dark line) and LDECA (solid (red) dots) results for the conductance $G/G_0$ as a 
function of $z$ (where $G_0=2e^2/h$ is the quantum of conductance), which can be compared to Fig.~4(a) in Ref.~\onlinecite{Bork11}. Since LDECA is exact at the Fermi energy \cite{Anda08}, the excellent agreement between SBMFA and LDECA gives support to the SBMFA results shown in this paper. The main difference with the experimental results is that our $G/G_0$ values increase more smoothly as z decreases. This can be associated to the fact that, as expected, the experimental results are very dependent upon the distance between the atoms.  As stated by Bork {\it et al.}, `mechanical relaxation' 
effects should be at play when a transition from `tunneling' to `point contact' occurs, as the tip gets closer to 
the surface. \cite{Bork11} Hence, at this transition the real distance $z$ and the associated hopping parameters $t_{\alpha \beta}$ and $t_{LR}$ are difficult to determine. Besides, other matrix elements such as $t_L$ or $t_R$, assumed to be constant, may also vary at this transition.

In the SBMFA, when the system is in the Kondo regime, there is a plateau at the Fermi energy ($E_{\rm F}=0$) in the 
renormalized energy level of each impurity ($\tilde \epsilon_{\alpha}= \tilde \epsilon_{\beta} = \tilde \epsilon$), 
as a function of gate voltage $V_g$ \cite{sbmfa,note3}. As can be seen in the inset of Fig.~\ref{fig3}, for the region 
of interest, the plateau is perfectly defined (compare (black) solid curve for $z=200$ with the almost identical (red) 
diamonds curve for $z=-60$), indicating that the system remains in the Kondo regime, although the splitting in dC is 
relatively large for the lowest values of $z$, as shown in Fig.~\ref{fig4}, discussed below. For larger values of 
$t_{\alpha \beta}$ and $t_{LR}$, the plateau is partially eliminated as shown in the (green) dashed curve for $z=-150$. 
Thus, for large hopping values there is an effective antiferromagnetic spin-spin correlation between the Co atoms 
that starts to suppress the Kondo regime, and the system enters a crossover region that is compatible with measurements 
presented in Fig.~S.6 of Ref.~\onlinecite{Bork11}. In order to confirm our SBMFA results, spin-spin correlations were calculated 
with LDECA. These are shown in the upper inset of Fig.~\ref{fig4}. It can be 
seen that even when a splitting is present, the AF spin-spin correlation between each impurity and its adjacent reservoir 
(calculated between the impurity and the first site of the adjacent non-interacting chain), 
which may be used to characterize the Kondo effect, is dominant in respect to the AF correlation 
between the impurities. This result coincides with the scenario provided by the SBMFA approach: as the system stays 
in the Kondo regime, the splitting in dC is a consequence of the hybridization of the Kondo resonances associated 
to each impurity \cite{Ribeiro12}.

The main panel in Fig.~\ref{fig4} shows a comparison, as a function of $z$, of the SBMFA 
dC-splitting (distance between peaks) and the experimental results from Ref.~\onlinecite{Bork11}, both 
in units of their corresponding single-impurity Kondo temperature $T_K^0$ ($75~K \approx 6.5~meV$ in 
the experiments). The SBMFA splitting is almost linear in $z$. In the lower inset we compare experimental and SBMFA 
results for the width of the dC Fano antiresonance in the interval $300 \geq z \geq 0$, scaled by $T_K^0$. We fit the 
SBMFA  dC curve with a Fano antiresonance to extract its width. The sharper decrease in the dip width 
for the experimental results at $z \approx 25$ can be ascribed to the relaxation process mentioned above. 
We stress the fact that as the magnitude of the splitting may change when the experiment is repeated, a 
qualitative and semiquantitative description of the experimental results should be satisfactory.

\section{Single impurity case}

\begin{figure}[ht]
\includegraphics[scale=0.3]{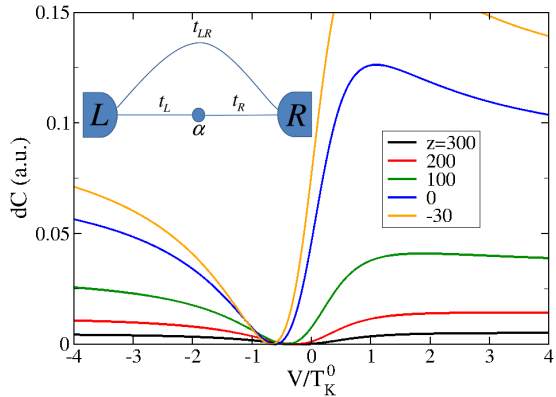}
\caption{Differential conductance for a single impurity as a function of $V/T_K^0$.  The behavior 
is markedly different from the case of two atoms, as in the experiments.}
\label{fig5}
\end{figure}

In the case of a single Co impurity, the experimental dC 
shows only one dip, which neither diminishes its width nor transforms into a peak as the distance 
between tip and surface is reduced \cite{Madhavan, Bork11}. In order to study the differences between 
the single- and double-impurity cases, we have carried out a $z$ dependence study for the single-impurity 
model. The model is depicted in the inset of Fig.~\ref{fig5}. The hopping  between the impurity and one of 
the reservoirs ($t_R$), modeling the STM tip) and the hopping between the two reservoirs ($t_{LR}$) are varied in the same way as 
in the double-atom case. The SBMFA results (main panel of Fig.~\ref{fig5}) show an asymmetric anti-resonance, 
as previously obtained \cite{Morr}. We checked that the dip's width does not decrease by changing $z$. 
Besides, the dip reaches its minimum at negative values of bias voltage, as noted in Ref.~\onlinecite{Morr} 
for an $S=1/2$ impurity, which is our case. The value of $G/G_0$ at the Fermi level (not shown) is much 
smaller than in the two-impurity case. Therefore, these results show that the reduction of the anti-resonance 
width, the appearance of a peak and its splitting, as was observed in the experiments of Ref.~\onlinecite{Bork11}, 
are a consequence of the presence of a second impurity, interacting with the first. 

\section{Conclusions}

Summarizing, the model proposed in this paper to study the transport properties through two Co 
atoms in series correctly describes the behavior observed in STM experiments for all the parameter range. In 
that respect, the inclusion of a direct hopping between the atoms, {\it besides} the one between the electron 
reservoirs, proves to be an essential ingredient. In addition, we find that the electrons interfere constructively 
along the two possible paths, but in the case of a single impurity these two same paths give completely different 
transport properties, as observed in the experiments. In our model, the direct and indirect couplings between the 
impurities result in an antiferromagnetic spin-spin correlation between them. This interaction is not strong enough to 
take the system out of the Kondo regime. The splitting in the differential conductance is indeed a consequence of the 
hybridization of the two Kondo resonances.  

\begin{acknowledgments}

We gratefully acknowledge fruitful discussions with C.~A. B\"usser, L.~O. Manuel, A.~E. Trumper, C.~J. Gazza, and E. Vernek. 
G.B.M. acknowledges financial support from NSF under Grants No. DMR-1107994, DMR-0710529, and MRI-0922811. E.V.A. acknowledges financial 
support from the Brazilian agencies CNPq (CIAM project) and FAPERJ (CNE).

\end{acknowledgments}


\begin{thebibliography}{99}
\bibitem{note2a} Note that this phase difference should not be confused with the one between 
even and odd channels in the usual two-impurity model, where $t_{LR}=0$.
\bibitem{note3} As a gate potential is not an experimental feature in Ref.~\onlinecite{Bork11}, we do not add 
a gate potential term to the Hamiltonian. The gate potential simply changes the energy level of each 
impurity $\epsilon_i=\epsilon_i+V_g$.
\end{thebibliography}
\end{document}